\documentclass[a4paper]{article}
\usepackage[a4paper, total={6in, 8in}]{geometry}
\usepackage[utf8]{inputenc}
\usepackage[english]{babel}
\usepackage[round]{natbib}
\usepackage{cite}
\usepackage{bigints}
\usepackage{multirow}
\usepackage{enumitem}  
\usepackage{graphicx}
\usepackage{amssymb}
\usepackage{subfig}
\usepackage{authblk}

\title{Convolution of Scale Invariant Continuous Ranked Probability Scores for Testing Experts' Statistical Accuracy}

\author[1]{Tina Nane}
\author[1,2]{Roger Cooke}
\affil[1]{Department of Applied Mathematics, Delft University of Technology}
\affil[2]{Resources for the Future (ret)}
\date{}

\begin{document}

\maketitle
\begin{abstract}
Computable solutions for expectations of Continuous Ranked Probability Scores are presented. After deriving a scale invariant version of these scores, a closed form for the convolutions of scores is presented. This closed form enables the testing experts' statistical accuracy. Results are compared with tests using a familiar Chi-square goodness of fit test using a recent data set of 6,761 expert probabilistic forecasts for which true values are known.

{\bf Keywords}  expert judgment, scoring rules, continuous ranked probability score, probability interval score, geometric probability, classical model, overconfidence, location bias.
\end{abstract}

\section{Introduction}
Scoring rules were introduced by de Finetti in 1937 as tools for encouraging honesty in eliciting subjective probabilities \citep{de1937prevision} and have been further developed by many authors including \citet{shuford1966admissible,savage1971elicitation, murphy1977value, Brown74, degroot1983comparison, gneiting2007strictly}. The latter reference gives an extensive overview. An expert receives a score as a function of his/her probability assessment and the realization. The score is strictly proper if the expert maximizes (for negatively sensed rules, minimizes) his/her expected score per item by, and only by, stating his/her true belief. Using a result of \citet{murphy1977value}, \citet{degroot1983comparison} gave an additive decomposition of strictly proper rules into ‘calibration’ and ‘refinement’ terms, thereby replacing Murphy’s ‘resolution’ (refinement applies only to well-calibrated experts). In the case of the logarithmic rule, refinement becomes Kullback-Leibler divergence of the sample distribution of realizations. Some authors \citep{Hersbach2000} adopt a framework in which nature picks a distribution for an unknown quantity and forecasters attempt to predict this distribution. Scoring rules for sets of variables play a key role in the classical model ({\em CM}, \citet{cooke1991experts}) for combining expert judgments.

Scoring rules for individual variables were not designed for evaluating or combining experts and are not generally fit for that purpose. Indeed, rewarding honesty is not the same as rewarding quality. 
A simple example illustrates this difference: Consider $100$ fair coin tosses. An expert assesses the probability of {\em heads} on each toss as $1/2$. With the standard scoring rules, the score for the outcome {\em heads} is the same as their score for {\em tails} on each toss. If the score for all $100$ assessments is a function of their $100$ scores for the individual tosses, then their score for $100$ tosses is independent of the outcome sequence; the outcome of $100$ {\em heads} receives the same score as $50$ {\em heads} and $50$ {\em tails}. Equal scores do not imply equal quality.

Another example concerns the quadratic rule for ‘rain / no rain’ events. This rule is positively sensed on  $[-1, 1]$ and assigns the quadratic score $2r – r^2– (1– r)^2$  if rain occurs, where $r$ is the expert’s probability of rain. Interchange $r$ and $(1– r)$ in case it does not rain. Consider 1000 next day forecasts of rain by two experts. Suppose the experts bin their forecasts as shown below \citep{cooke2014validating}.

\begin{table}[ht]
    \centering
    \small
    \begin{tabular}{|c|c|c|c|c|c|c|c|c|c|c|c|c|}
    \hline
       \multicolumn{2}{|c|}{Probability of rain next day} & $5\%$ & $15\%$ & $25\%$ & $35\%$ & $45\%$ & $55\%$ & $65\%$ & $75\%$ & 
       $85\%$ & $95\%$ & Totals \\
       \hline
        \multirow{2}{*}{expert 1}  & assessed & 100 & 100 & 100 & 100 & 100 & 100 & 100 & 100 & 100 & 100 & 1000 \\
         \cline{2-13}
         & realized & 5 & 15 & 25 & 35 & 45 & 55 & 65 & 75 & 85 & 95 & 500 \\
         \hline
         \multirow{2}{*}{expert 2}  & assessed & 100 & 100 & 100 & 100 & 100 & 100 & 100 & 100 & 100 & 100 & 1000 \\
         \cline{2-13}
         & realized & 0 & 0 & 0 & 0 & 0 & 100 & 100 & 100 & 100 & 100 & 500 \\
         \hline
    \end{tabular}
    \caption{Binned subjective probability of rain assessments for the 1000 next days by two experts.}
    \label{tab:my_label}
\end{table}
  
Ten probability bins are considered, each associated with a forecast probability of rain. The experts' assessments are equally informative in the sense that they assign the same probabilities to the same number of days. Expert 1 is statistically perfectly accurate, that is, the empirical frequency of actual rainy days from the assessed 100 days is identical with the probability associated with each bin. Expert 2 is massively  inaccurate statistically. The sample distributions bear little resemblance to his/her assessed probabilities $(5\%, \ldots, 95\%)$. Expert 1 has an average quadratic score of 0.67 and Expert 2 an average quadratic score of 0.84. Expert 2 gets a better quadratic score because the higher resolution in the sample out-weighs statistical inaccuracy. Such examples make it difficult to explain intuitively what the scores 0.67 and 0.84 mean. For more discussion, see  \citep{ cooke1991experts,  cooke2014validating}. In the context of expert judgment we would like to reward both honesty {\em and} quality with scoring rules that are intuitive and easily explained. This requires numerical insight into the rules’ behavior. 

The negatively sensed Probability Interval Score ($PIS$) and its related Continuous Ranked Probability Scores ($CRPS$)  have recently been applied to COVID-19 probabilistic predictions \citep{ray2020ensemble}. \citet{gneiting2007strictly} note: “Applications of the $CRPS$ have been hampered by a lack of readily computable solutions to the integral (\ref{crps})" (see below). 

This article presents computable solutions to this integral which then allow us to study its trade offs between statistical accuracy and "sharpness". The $CRPS$ can be thrown into a scale invariant form which offers significant advantages. After reviewing the $PIS$ and $CRPS$, we introduce a re-parametrization of $CRPS$ by transforming the realizations according to the probability integral transformation of an expert’s assessed cumulative distribution function ($CDF$).  An expert with $CDF\,\, F_X$ for continuous variable $X$ is scored not with respect to the realization $y$ but with $F_X(y)$, the quantile of the distribution of $X$ realized by $y$.  The proposed CRPS transformation has several advantages: 
\begin{enumerate}[label=(\roman*)]
    \item The transformed $CRPS$ becomes scale invariant.
    \item The expert’s sampling distribution of transformed $CRPS$ can be expressed in closed form
    \item The density of convolutions of transformed $CRPS$ scores for independent variables is available in closed form.
    \item Transformed $CRPS$ can then be used to test the expert’s statistical accuracy without recourse to an asymptotic distribution.
\end{enumerate}

On the downside, $CRPS$ is insensitive to location bias. If the experts assess only certain quantiles, a second downside is that continuous $CDFs$ must be interpolated before applying $CDFs$. 

After introducing  the Probability Interval Score and the Continuous Ranked Probability Score,  computable examples of the latter are given. This motivates  a scale invariant version of CRPS to be used in testing experts' "statistical accuracy", a term denoting goodness-of-fit tests adapted to expert judgments. The closed form of the convolution of scale invariant $CRPS$ scores is introduced. Using a recently compiled database of expert judgments with realizations \citep{cooke2021expert}, the results of this test are compared with the statistical accuracy score of $CM$. The scores are also compared with regard to rewarding proximity of the medians (considered as  point forecasts) to the realizations. We conclude that the scale invariant $CRPS$ better rewards proximity to the median while failing to punish location bias. The closed form convolution also confers advantages with respect to $CM$.   

\section{Probability Interval Scores}
Numerical insight into the behavior of these scores requires a bit of effort.  For the $(1 - \alpha)$ uncertainty interval $[L, U]$, with upper (lower) bound $U (L)$, the $PIS$ (negatively sensed) \citep{Aitchison&Dunsmore68} for realization $y$ is

\[
(U-L) + \frac{2}{\alpha}\times [(L-y)_+ + (y-U)_+]
\]
where $X_+ = X$ if $X > 0$ and $X_+= 0$ otherwise. Note that $s = 2/\alpha$ is the slope of the overconfidence penalty for $y \notin[L,U]$. The length $\|U-L\|$ is called the ``sharpness''; small values reward concentrated probability mass. \\

To better understand the characteristics of $PIS$, consider $Y\sim U[0,1]$ and the  $(1 - \alpha)$ uncertainty interval $[L, U]$. Then

\[
E_Y[PIS(y)]= U-L +\frac{2}{\alpha} \int_0^L (L-x)\,\mathrm{d}x +\frac{2}{\alpha} \int_U^1 (x-U)\,\mathrm{d}x = U-L +\frac{1}{\alpha}\left[L^2+(U-1)^2\right].
\]

For the central $0.9$ interval $[0.05, 0.95]$, the expected PIS is $0.95$. Suppose an expert prefers to give an $80\% $ interval$, [0.1, 0.9]$, then the expected score is $0.9$. This is better than $0.95$ because the prediction interval is sharper. An expert seeking to optimize (i.e., minimize) his/her expected score might take a central $2\%$ prediction interval $[0.49, 0.51]$ with expected score of $0.51$. The way in which the $PIS$ trades overconfidence for sharpness may strike some as counter-intuitive. For example, an expert claiming that the degenerate interval $[0.5, 0.5]$ has $40\%$ probability of catching the realization would achieve an expected score of $0.833$, better than the score of the $90\%$ central interval. The sharpness of an interval of zero length outweighs the overconfidence of claiming $40\%$ mass at the point $0.5$.  

\section{Continuous Ranked Probability Score}
Consider $y$ an unknown scalar quantity of interest. Suppose $y$ has a true forecast cumulative distribution function ($CDF$) $F_Y$, characterizing the distribution of a random variable $Y$, which is not  known. An expert provides $CDF \,\,F_X$ which (s)he believes to be the distribution of $Y$.
We assume both $F_Y$ and $F_X$ are  continuous and strictly increasing on their support. The continuous ranked probability score (CRPS) is defined as \citep{Brown74}  

\begin{equation}
CRPS(F_X,y)=\int_{-\infty}^\infty [F_X(x)-  1_{\{x\geq y\}}]^2dx.
\label{crps}
\end{equation}

Lower values indicate better performance. CRPS is known to be strictly proper relative to a class of Borel probability measures with finite first moment \citep{gneiting2007strictly}. As  mentioned in the introduction, lack of  readily computable solutions for \eqref{crps} have restricted the use of CRPS score. To understand the behavior of the CRPS score, let us consider $X\sim U[L,H]$, with $0<L<H<1$. As we will show later, this particular choice of distribution is relevant for the development of our proposed score. For  $y\in[0,1]$: 

\begin{equation}
CRPS(F_X,y)=\left\{ \begin{array}{c}
 \begin{array}{ccc}
L-y+\frac{H-L}{3} &  for\ 0\leq y<L \vspace{0.5mm} \\ 
\frac{(y-L)^3}{3(H-L)^2}-\frac{(y-H)^3}{3(H-L)^2} & for\ y\in[L,H] \vspace{0.5mm} \\
y - H + \frac{H-L}{3} & for\ H<y\leq 1.
\end{array}
 \end{array}
\right.
\end{equation}

\begin{figure}[hbt!]
\subfloat[ L=0.3, H=0.4 \label{fig:test1}]
  {\includegraphics[width=.3\linewidth]{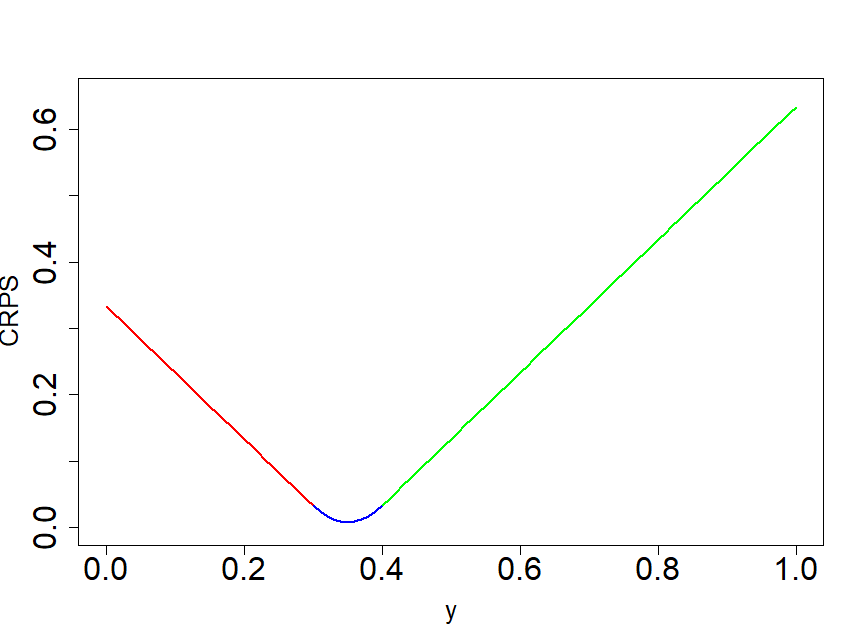}}\hfill
\subfloat[L=0.6, H=0.7 \label{fig:test2}]
  {\includegraphics[width=.3\linewidth]{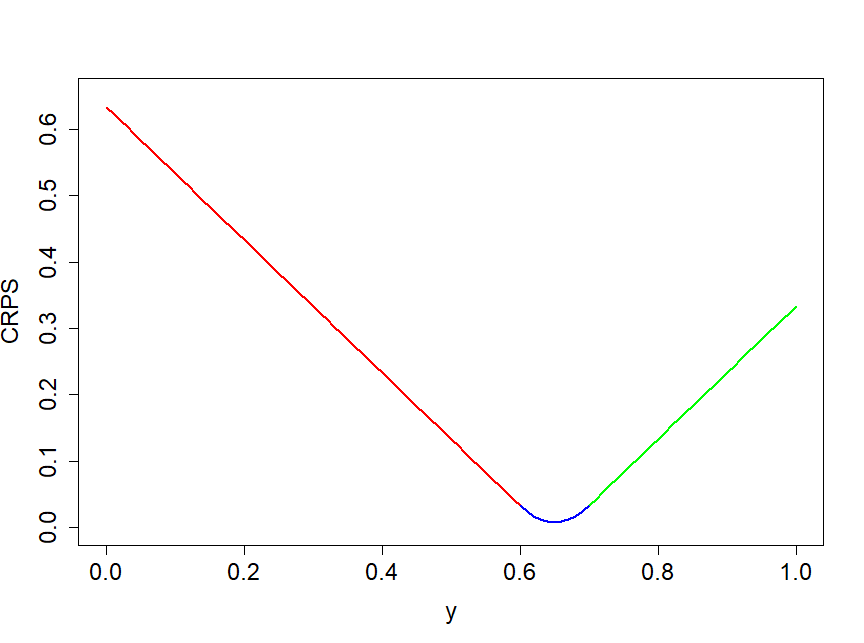}}\hfill
\subfloat[ L=0.3, H=0.7 \label{fig:test3}]
  {\includegraphics[width=.3\linewidth]{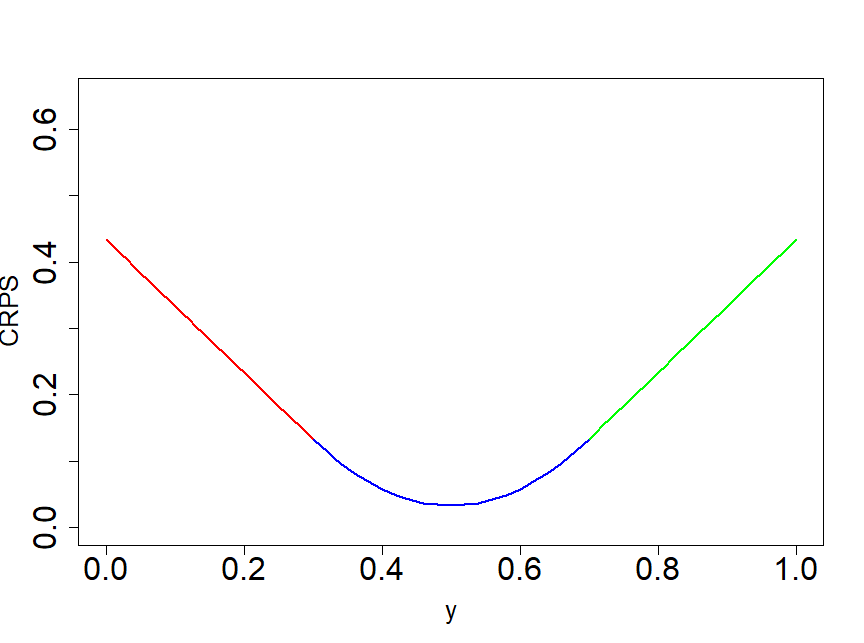}}
\caption{CRPS score for $X\sim U[L,H]$ and $y\in[0,1]$. Differences for $y<L$ (red), for~$y\in[L,U]$(blue), and $y>U$(green) are highlighted.}
\label{fig:crps}
\end{figure}

Figure \ref{fig:crps} shows the CRPS score as a function of $y$, for different cases of $L$ and $H$. Cases when $y$ falls within and outside the $F_X$ support are highlighted. 


The expectation of the CRPS score, which may be infinite, is given by

\begin{equation}
E_Y[ CRPS(F_X,Y)] = \int_y \int_x [F_X(x) – 1_{\{x\geq y\}}]^2 \mathrm{d}x \,\mathrm{d}F_Y(y). 
\label{expected_crps}
\end{equation}

We discuss some computable solutions for this expectation.
 
\subsection{Computable solutions}
\begin{figure} [hbt!]
\centering
\includegraphics[width=0.8\linewidth]{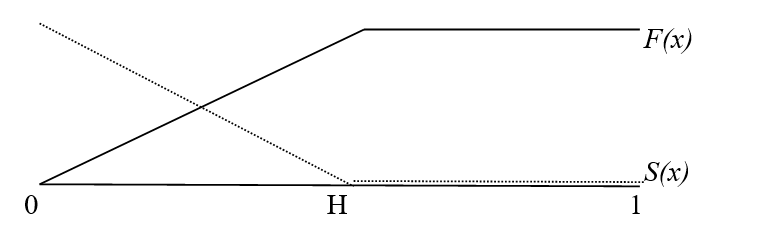}
\caption{Cumulative distribution and survivor function of an uniformly distributed random variable on [0,H].} \label{fig1}
\end{figure}

Consider  $Y \sim U[0,1]$ and an assessment of $Y$'s distribution by an expert as that of random variable $X$,  $X \sim U[0, H],\, 0 < H < 1$.  The expert thinks values greater than $H$ are impossible, although these can in fact arise.  The expected CRPS is computed based on the distribution of $Y$. The $CDF$ of $X, F(x) = x/H$, for $x\in[0,H]$ and $F_X(x) = 1,\, $for $ x\geq H$, along with the survivor function of $X, S(x) = 1 - F(x)$ are shown in Figure \ref{fig1} (see also \citet{Candille&Talagrand2005}).

Then  $E_Y[ CRPS(F_X,Y)] = \int_0^1  \int_0^1 [F(x) - 1_{x\geq y}]^2 \mathrm{d}x \, \mathrm{d}y$ is computed in 2 steps:\\

\noindent A) For $y < H:$
 \begin{equation*}
  	\int_0^H  \left[ \int_0^y \frac{x^2}{H^2}  dx + \int_y^H \left(\frac{H-x}{H}\right)^2 dx \right] dy =  \int_0^H  \left[\frac{y^3}{3H^2}  + \int_{H-y}^0  \frac{z^2}{H^2} (-dz) \right] dy  = \frac{H^2}{6}.
\end{equation*}
B) For	$y >H:$

\begin{align*}
  \int_H^1 \left( \int_0^H \frac{x^2}{H^2} dx+ \int_H^y\, dx + \int_y^1 0 \, dx\right) dy &=  \int_H^1 \left( \frac{H}{3} + y-H\right)\, dy = \frac{H(1-H)}{3} + \int_0^{1-H} \, z dz \\
&= \frac{H(1–H)}{3} + \frac{(1–H)^2}{2}.
\end{align*}

Therefore:

\begin{equation}
E_Y[CRPS(F_X,Yy)] = \frac{H^2}{6} + \frac{H(1–H)}{3} + \frac{(1–H)^2}{2}.   
\label{eq3}
\end{equation}

As noted by \citet{Hersbach2000}, these results acquire a physical dimension. The result for (A) is the score an expert with $X \sim U[0,H]$ expects, namely $H^2/6$, which has the physical dimension of $H^2$.  If $X$  is in meters and changes to centimeters, the expected score increases by a factor $10^4$. 
\vspace{.5cm}

If  $X \sim U[L, H]$ , for $0 <  L < H < 1$, then the same method of calculation applies mutatis mutandis. If $L = 1-H$, with  $H \geq 0.5$, then the contributions from $x <  y$  and $y <  x$ are equal and we need only to double the contribution from $x < y$. If  $y < L$, the contribution from $x<y$ is zero.  We compute
\begin{align*}
\int_L^H  \int_L^y  F(x)^2\,dx\, dy  + \int_H^1 \int_L^y  F(x)^2\,dx\, dy  &= \\
 \int_L^H  \int_L^y  \frac{(x–L)^2}{(H–L)^2}\,dx\, dy + \int_H^1 \left[\int_L^H \frac{(x-L)^2}{(H-L)^2}\,dx + \int_H^y\,  dx \right]\,dy &= \\
\int_L^H \frac{y^3}{3(H-L)^2} \,dx + \int_H^1 \left[\frac{(H-L)}{3} +(y-H)\right] \,dy &= \\ 
\frac{(H-L)^2}{12} + \frac{(H-L)(1-H)}{3} + \frac{(1-H)^2}{2}.
\end{align*}
Adding the identical contribution from $y < x$ gives:
\begin{equation}
E_Y[CRPS(F,Y)] = \frac{(H-L)^2}{6} +\frac{2(H-L)(1-H)}{3} + (1-H)^2.
\label{eq4}
\end{equation}

Again, the score inherits a physical dimension from $X$. Substituting $L = 1- H$  in (\ref{eq4}) we find that $E_Y[CRPS(F,Y)]$ for  $X \sim U[1 - H, H]$ is equal to $E_Y[CRPS(F,Y)]$ for $X \sim U[0, H]$ from (\ref{eq3}). This holds for any $0<L<H<1$, with $L=1-H$. Hence, for  $X \sim U[0, 0.7],\,\, E_Y[CRPS(F,Y)] = 0.1966  = E_Y[CRPS(\tilde{F},Y)]$, for $ \tilde{X} \sim U[0.3, 0.7]$. By the same token, $X \sim U[0,\,0.5]$ yields the same expected score of $1/4$ as $\tilde{X}$ with distribution $\delta(0.5)$ concentrated at $0.5$. 

If $F_i \rightarrow F$, then $E_Y[CRPS(F_i,y)]  \rightarrow E_Y[CRPS(F,y)]$, by the Helly-Bray theorem \citep{billingsley2013convergence}. It follows that, if $F_n \rightarrow U[0,\,0.5]$ and $\tilde{F}_n  \rightarrow \delta(0.5)$, then for all $\varepsilon > 0$ and for all sufficiently large $n$, $|E_Y[CPRS(F_n,Y)] - E_Y[CPRS(\tilde{F}_n,Y)]| < \varepsilon$. This illustrates how the $CRPS$ compensates loss of statistical accuracy by a gain in “sharpness”, and again illustrates that equal scores do not entail equal quality. 

Note that if the probabilistic forecast is the distribution of $Y$ (uniform $[0,1]$), then the expected $CRPS$ score is $1/6$, by \eqref{eq3}. For $0 < H < 1$, $E_Y[CRPS(F,Y)] > \frac{1}{6}$, by \eqref{eq3} and for $0.5 < H < 1, \, L=1-H$, $E_Y[CRPS(F,Y)]> \frac{1}{6}$, by \eqref{eq4}. An expert would receive a better (lower) expected score if their probabilistic forecast were equal to the distribution of $Y$.

\section{Test for statistical accuracy}\label{TSA}

We would like to test the  hypothesis that $Y$ follows the expert's assessed distribution $F_X$. Applying the probability integral transformation, let $U=F_X(X)$ and define $V = F_X(Y)$. Then $F_U(u) = u$. The hypothesis that $F_X = F_Y$ is equivalent to the hypothesis 
\[
H_0:\,\,\,  V \sim\, U[0,\,1].
\]
\vspace{0.5mm}
In this case  $CRPS$ can be written, for the realization $v$ and for $U$ uniformly distributed on $[0,1]$, as  

\begin{align}
CRPS(F_U,v)& =\int_{-\infty}^\infty [u-1_{\{u\geq v\}}]^2\mathrm{d}{u} \nonumber \\
    & = \int_{-\infty}^0 (0-0)^2\mathrm{d}{u}+\int_0^v u^2\mathrm{d}{u} +\int_v^1 (u-1)^2\mathrm{d}{u} \nonumber
    +\int_1^\infty (1-1)^2\mathrm{d}{u}\\ \label{crps_unif}
    & = \frac{v^3}{3} - \frac{(v-1)^3}{3}.
\end{align}

The range of the $CRPS(F_U,v)$ is $\left[ \frac{1}{12},\frac{1}{3} \right]$, for $v\in[0,1]$ and $U\sim U[0,1]$. The distribution of $CRPS$ is the distribution of  the random variable
\vspace{.5mm}
$$
\frac{1}{3} \left[V^3 -(V - 1)^3\right]  = \frac{1}{3} - V + V^2 ,
$$ 
\vspace{.5mm}
taking values in $[\frac{1}{12},\frac{1}{3}] \,\,$(lower values are better). Under the null hypothesis, $V$ is uniform $[0,\,1]$. For  fixed $Q \in [\frac{1}{12},\frac{1}{3}]$, to find the probability that $CRPS(F_U,v) \leq Q$, under the null hypothesis, we find the roots of $V^2 - V +(\frac{1}{3}  - Q) = 0:$
\vspace{.5mm}
\[
v_{1,2} = \frac{1 \pm \sqrt{1 - 4(\frac{1}{3} - Q)}}{2} = \frac{1 \pm \sqrt{4Q –\frac{1}{3}}}{2}.
\]

Collecting the mass between the two roots, we obtain the CDF
\[
P(CRPS(F_U,v)\leq x) = \sqrt{4x -\frac{1}{3}},
\]
with density  
\begin{equation}
\frac{2}{\sqrt{4x-\frac{1}{3}}},\,\,\, x \in \left(\frac{1}{12}, \, \frac{1}{3}\right).\label{dens}
\end{equation}

\begin{figure*}[hbt!]
\centering
\includegraphics[width=0.85\linewidth]{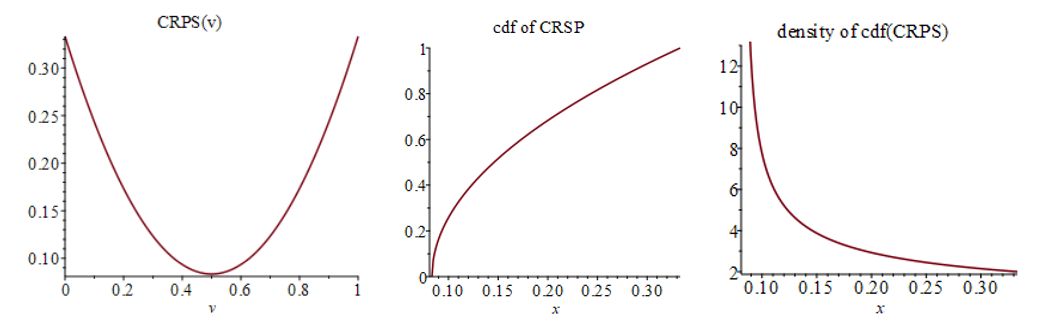}
\caption{CRPS function, for $U\sim U[0,1]$ and $v\in[0,1]$, together with its cumulative and density functions, under the null hypothesis $H_0$.}\label{dens&CDF}
\end{figure*}

Figure \ref{dens&CDF} shows $CRPS(F_U,v)$, for $v \in [0, 1]$ and $U\sim U[0,1]$, together with its $CDF$ and density under the null hypothesis $H_0$.

From Figure \ref{dens&CDF} it is evident that the score $CRPS(F_U,v)$ is symmetric around the value $v=0.5$. This is different from the behavior of $CRPS(F_U,v)$ exhibited in Figure \ref{fig:test1} or \ref{fig:test2}, and it illustrates a feature of the scale invariant version of $CRSP$.

From equation \eqref{dens}, we can easily compute

\begin{eqnarray}
\centering
E_V\left[CRPS(F_U,V)\right] &=& \frac{1}{6}\nonumber \\
E_V\left[CRPS^2(F_U,V)\right] &=& \frac{1}{16}\left(\frac{1}{5} + \frac{1}{3}\right) \nonumber \\
Var_V\left(CRPS(F_U,V)\right) &=& 0.005555\underline{5} \nonumber
\end{eqnarray}

It is handier to consider the following transformation 

\begin{equation}
Z(U,V)=4 CRPS(F_U,V)-\frac{1}{3}.
\label{transf}
\end{equation}

Then $Z$ has $CDF$ and density

\[
F_Z(x)=\sqrt{x}, \,\,\,\text{ and}\,\,\,\,\, f_Z(x)=\frac{1}{2\sqrt{x}},\,\,  x\in[0,1].
\]

Note that $f_Z$ is the density of $U^2$, where $U\sim U[0,1]$. So far, only one unknown scalar quantity of interest, and expert's resulting $CRPS$ score, have been considered. 

Suppose an expert provides uncertainty assessments for $n$ random variables. The emerging question is how to aggregate the $CRPS$ scores of each of the $n$ variables? The transformation \eqref{transf}, and the observation that the density $f_Z$ is the density of a squared uniform random variable are again handy. 

If we assume the $n$ variables to be independent, then we need to consider $Z_1,\ldots,Z_n$ independent variables, each with density $f_Z$. For these, we need to find the density of $Z^{(n)}=Z_1+\cdots+Z_n$. Or, in terms of the squared uniform random variables, we need to find the density of $S_n=U_1^2+\cdots+U_n^2$.

\citet{weissman2017sum} provides closed form distributions for $S_n$, for $n=3,4,5,6,8,10,12$ and their graphical representations. A connection is also made with a topic of geometrical probability, that is, finding the cumulative distribution function of $S_n$, $P(S_n\leq s)$ is equivalent to finding the volume of the intersection between the unit n-cube and the ball of radius $\sqrt{s}$, in $\mathbb{R}^n$, when both are centered at the origin. In his comment to \citet{weissman2017sum}, \citet{forrester2018comment} observes that the more generic volume problem posed by \citet{xu1996volume}, of finding the volume of the intersection of a cube and a ball in n-space has already been solved by B. Tibken and D. Constales (\citet{RousseauRuehr1997}). \citet{weissman2017sum} reports that Constales' solution to the volume problem involves a method based on Fourier series and implies that, for general $n$,

\begin{equation}
F_n(s)=P(S_n\leq s)= \frac{1}{6} +\frac{s}{n}+\frac{1}{\pi} \textbf{Im} \sum_{k=1}^\infty \left[ \left( \frac{C(2\sqrt{k/n})-i S(2\sqrt{k/n})}{2\sqrt{k/n}}\right)^n \frac{e^{2\pi i k s/n}}{k} \right],
\label{exact_distr}
\end{equation}
where $S(x)=\int_0^x sin (t^2)\,\mathrm{d}t$ and $C(x)=\int_0^x cos (t^2)\,\mathrm{d}t$ denote the Fresnel integrals and $\textbf{Im}$ is the imaginary part of a complex number. 

\begin{figure}[hbt!]%
    \centering
    \subfloat{{\includegraphics[width=6.2cm]{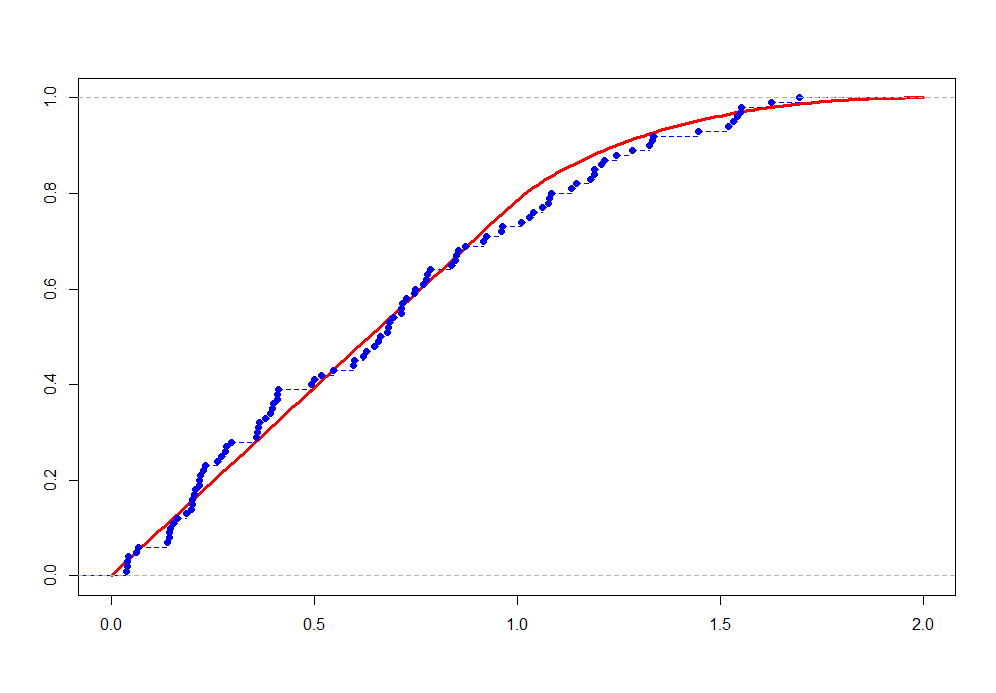} }}%
    \qquad
    \subfloat{{\includegraphics[width=6.2 cm]{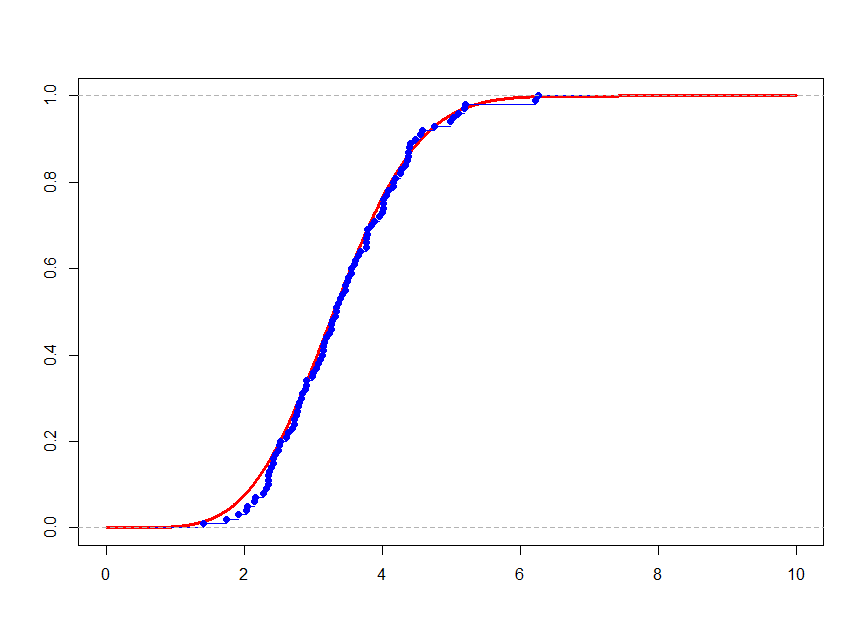} }}%
    \caption{The exact distribution (red) and empirical distribution function of the sum of squared simulated uniform observations (blue), for n=2 (left) and n=10 (right).}%
    \label{fig:example}%
\end{figure}

Figure \ref{fig:example} graphically compares the above cumulative distribution for n=2 (left) and n=10 (right) with the empirical distribution function of the corresponding sum of squared uniform observations. 100 observations were sampled for both empirical distribution functions. The cumulative distribution function was implemented in \verb!R!, by making use of functions implementing the Fresnel integrals in the \verb!pracma! package~\citep{borchers2022package}.

\section{Expert data}
In this paper, we use data from $49$ studies involving 526  experts assessing in total $580$ calibration variables from their fields for which realizations are known (four experts from the original data were dropped because they did not assess all calibration variables in their respective panels). In total there are 6,761 expert probabilistic forecasts of variables from their fields for which true values are known.  The data is described and referenced in \citep{cooke2021expert}. The supplementary information for that reference gives a description of the Classical Model, whose relevant aspects are briefly reviewed here.

\begin{figure*}[hbt!]
\centering
\includegraphics[width=0.65\linewidth]{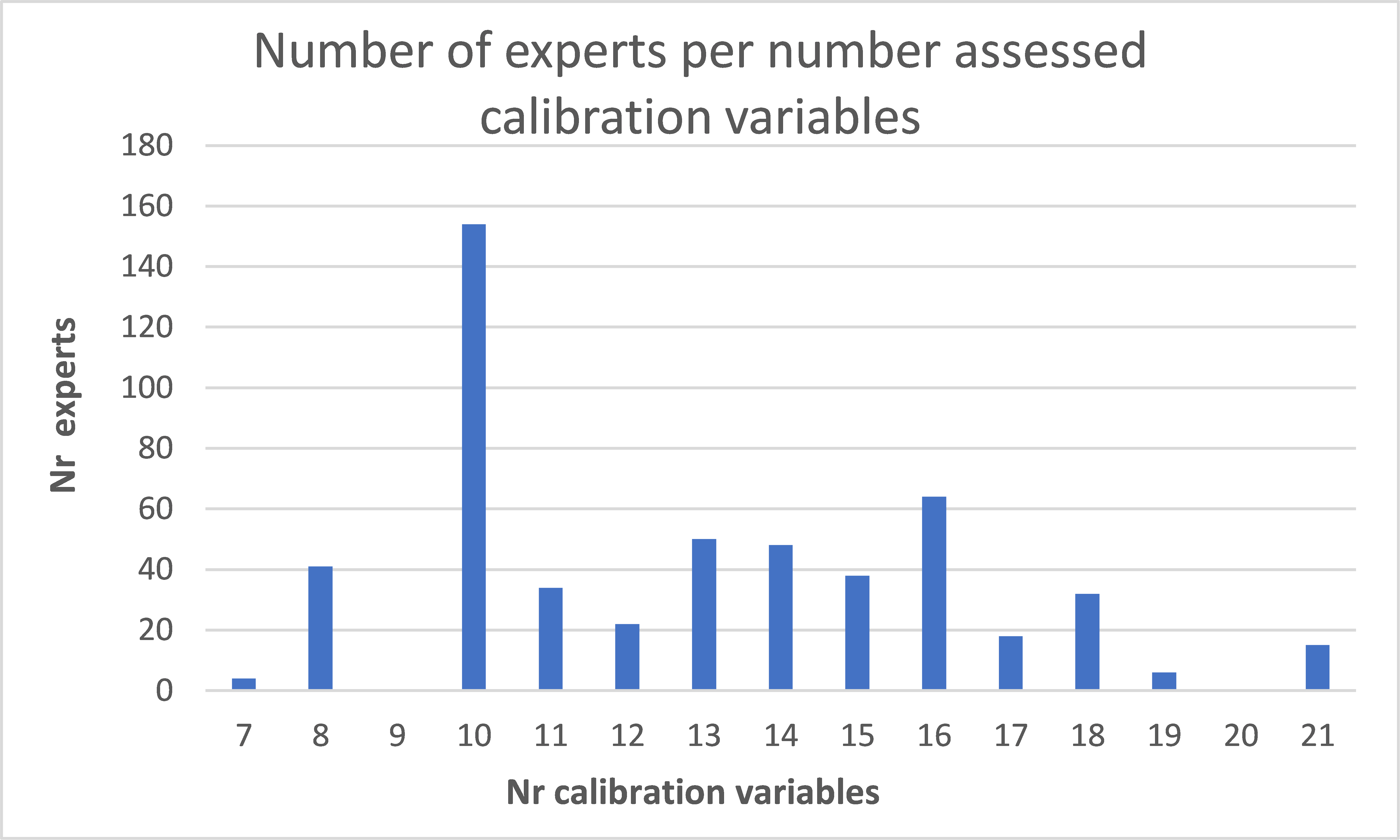}
\caption{Number of experts per number of assessed calibration variables.}\label{NrCal}
\end{figure*}

The number of assessed calibration variables differ per study and Figure(\ref{NrCal}) provides information about this. Experts assessed at least 7 and at most 21 calibration variables during all studies, and 10 calibration were used in 21 of the 49 studies. 

\section{Results of Statistical Tests}
This section  compares statistical accuracy ($SA$) tests based on the classical model $(CM ,\,$\citet{cooke1991experts})  versus based on $CPRS$. In the classical model, $SA$ is measured as the probability of falsely rejecting the hypothesis that a probabilistic assessor is statistically accurate. It is, in other words, the $P$-value of rejection for this hypothesis. We hasten to add that $CM$ does {\em not} test and reject expert hypotheses but, in compliance with proper scoring rule theory for sets of assessments, uses this $P$-value to measure the degree of correspondence between assessments and data in forming weighted combinations of expert distributions. 

In the data used for this analysis, expert assessments take the form of fixed quantiles, $5\%, \, 50\%,$, $95\%$, from the assessor's subjective distribution for a continuous unknown quantity. When $n$ true values for a number of such quantities are observed, we compute the sample distribution $s$ of inter-quantile relative frequencies and compare this with the theoretical inter-quantile mass function $p=(0.05, 0.45,0.45,0.05)$. The test statistic is $2nI(s,p)$, where $I$ is the Shannon relative information (log likelihood ratio) and $n$ is the number of calibration variables. Assuming that the realizations are independently sampled from the assessor's distributions, this statistic is asymptotically chi-square distributed with degrees of freedom equal the number of assessed quantiles. Thus, $SA$ is measured as $1-F_{\chi^2}(2nI(s,p))$ where  $F_{\chi^2}$ is the $CDF$ of a $\chi^2$ distribution with three degrees of freedom. High scores (near $1$) are good, low scores (near $0$) mean it is unlikely that the divergence between $s$ and $p$ should arise by chance. Note that $CM$ relies on an asymptotic approximation which for a small number of observations is not very good \citep{cooke2014validating}. Simulations for ten calibration variables are provided in \citep{hanea2021depth}. It is deemed capable of detecting only large differences in experts' performances. It uses only the assessed quantiles and does not rely on an interpolated $CDF$.

When applied to this expert judgment data, a test based on $CRPS$ must interpolate an expert's $CDF$. For this purpose we follow CM and adopt a minimally informative distribution relative to a uniform  support (chosen by the analyst{\footnote {In principle, any background measure supporting the experts' quantiles and realization may be used, this data imposes the uniform background for convenience}) which complies with the expert's quantile specification. The resulting CDF can be found in \citet{hanea2021depth}. $CM$ uses this interpolation for computing an expert's informativeness, but not for computing SA.\\

Consider  $n$ observations of continuous variables assessed by an expert $e$. The following procedure  calculates $SA$ for the $CRPS$ statistic

\begin{enumerate}
    \item For each  realization $y_i=1,\ldots,n$, compute $q_e^i= F_{i,e}(y_i)$, the quantile of $y_i$  in expert  $e$'s $CDF$ $F_{i,e}$.
    \item The $SA$ hypothesis $H_0$ entails that these quantiles are  independent samples from a uniform distribution. Under this hypothesis, $CRPS_i(e) = CRPS(F_U,q_e^i)$ can be computed from \eqref{crps_unif}. 
    \item For each $i=1,\ldots,n$, compute $z_i(e)=4CRPS_i(e)-1/3$, which is given in \eqref{transf}
    \item Compute $s(e)=F_n(\sum_{i=1}^n z_i(e))$, where $F_n$ is the exact distribution of the sum of $n$ independent squared uniform variables, given in \eqref{exact_distr}.
\end{enumerate}

Note that the procedure can be applied for continuous and invertible CDFs. $CRPS$ uses an exact instead of an asymptotic distribution for the convolution of these $CDF$s.  From Figure \ref{dens&CDF} it is evident that the score $CRPS(v)$ for value $v$ is symmetric around the value $0.5$. The distribution of the sum of such variables is insensitive to location bias in the following sense: the score for $2n$ observations of $0.4$ is the same as for $n$ observations of $0.4$ and $n$ observations of $0.6$. Figure \ref{dens&CDF} also shows that $CRPS$ is also insensitive to under-confidence: An expert whose probability transformed realizations are all $0.5$  scores better than one for whom the hypothesis of Section \ref{TSA} that $F_X = F_Y$ holds. Under-confidence is rare with expert judgment. Over-confidence, on the other hand, is not rare and the $CRPS$ score is sensitive to over-confidence (see Figure \ref{freq_percentiles}).

\begin{figure*}[hbt!]
\centering
\includegraphics[width=0.65\linewidth]{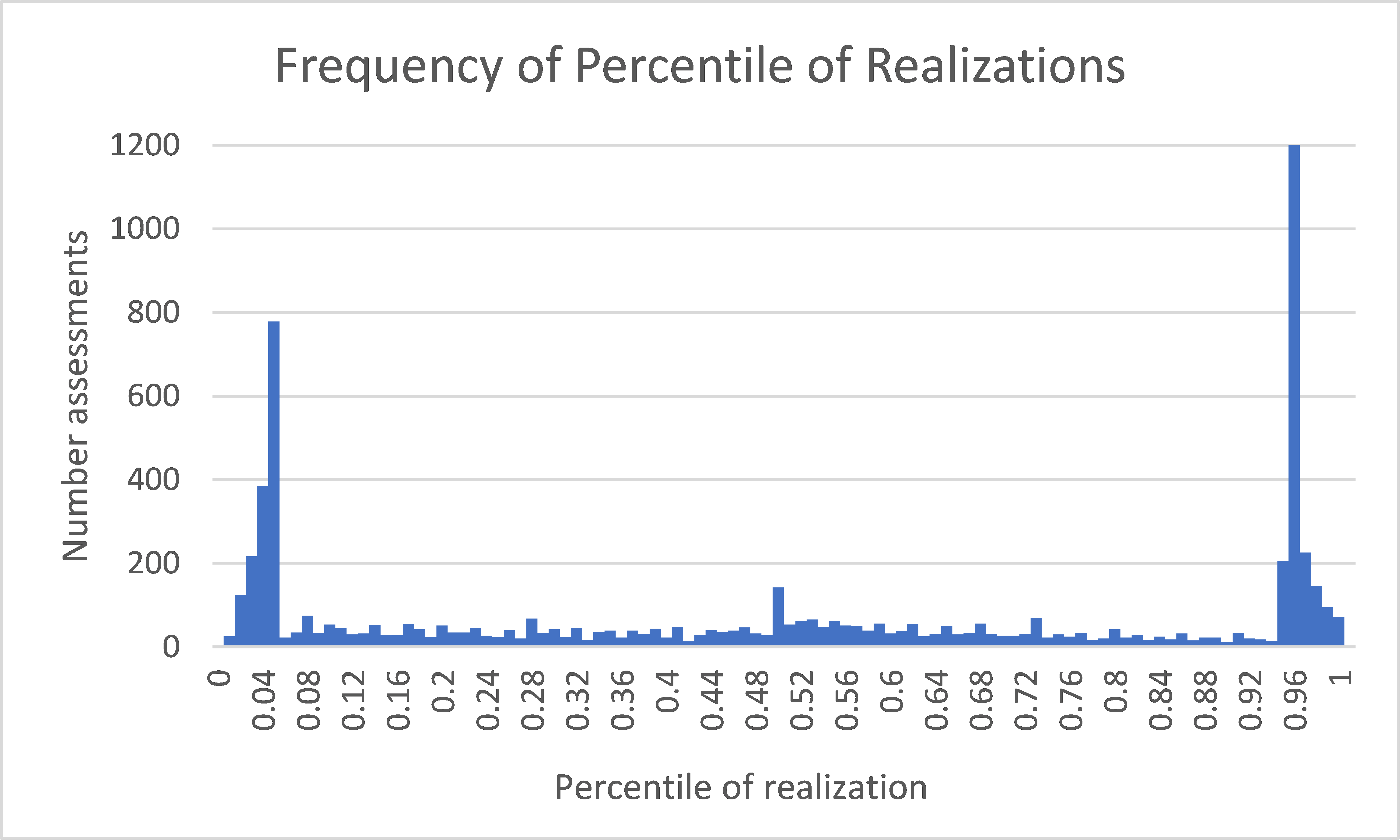}
\caption{Frequency of percentiles of realizations for $6761$ expert probabilistic forecasts from 49 studies.}\label{freq_percentiles}
\end{figure*}

Figure \ref{SA_CM_CRPS} plots the SA scores for $526$ experts based on $CM$ and on $CRPS$. Although the drift of the two scores is similar, there is substantial scatter. $CRPS$'s log geomean $SA$ score is $-2.76$ while that of $CM$ is $-3.47$; in this sense $CRPS$ is less severe.

\begin{figure*}[hbt!]
\centering
\includegraphics[width=0.65\linewidth]{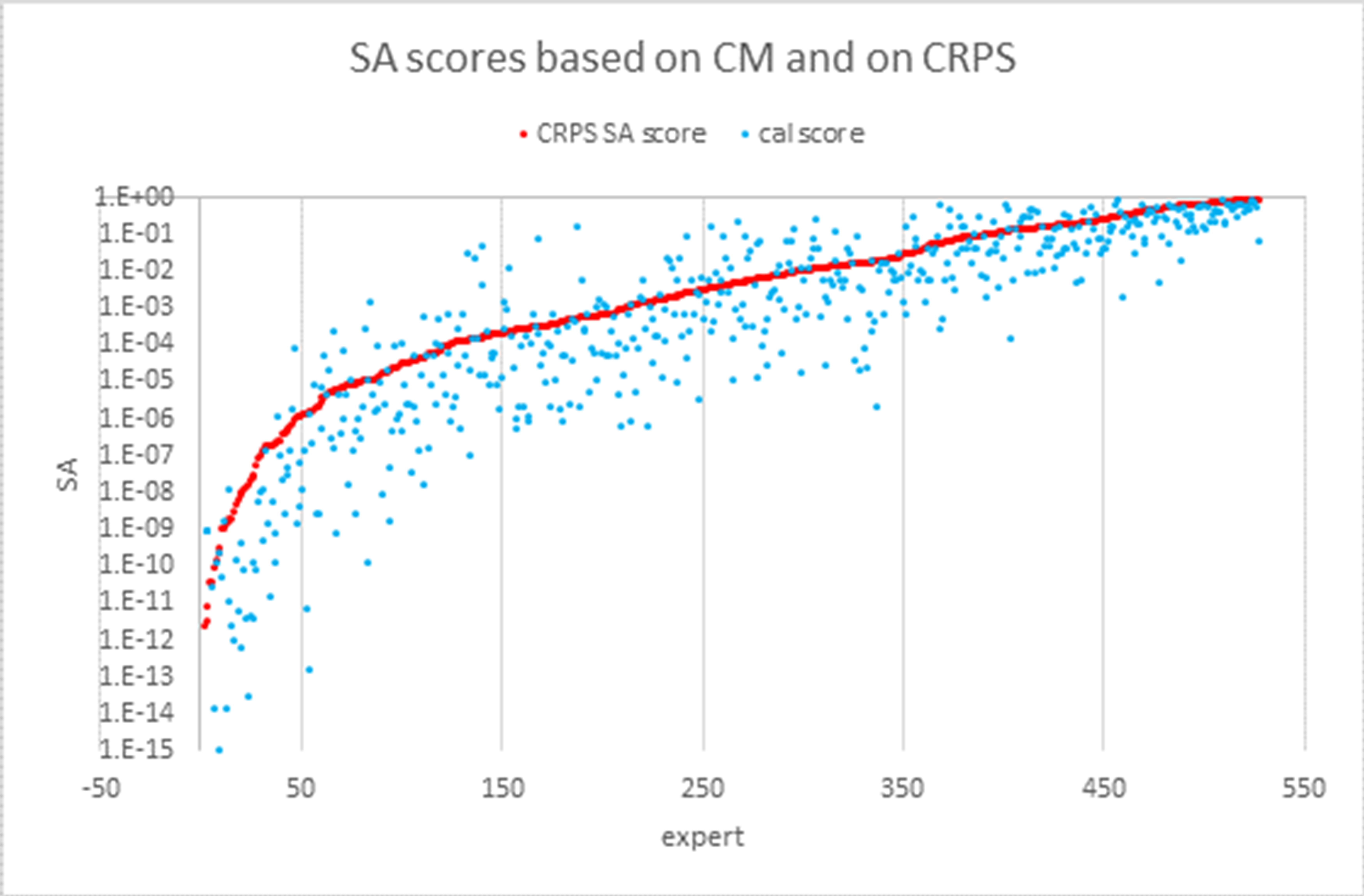}
\caption{Statistical accuracy of 526 experts with respect to CRPS (red) and CM (blue). SA scores are ordered by CRPS SA scores.}
\label{SA_CM_CRPS}
\end{figure*}

We define an expert's location bias as the absolute difference between the percent of realizations above the medians and $50\%$. Location bias of 50\% means that all realizations are above or all realizations are below the medians. Figure \ref{SA_CM_CRPS_LB} black circles the $CM$ scores of those experts for whom the location bias is greater or equal to $20\%$. $CM\,SA$ scores of experts for whom the location bias is $50\%$ ({\em all} realizations were either below or {\em all} above the medians) are red circled. The location bias of these circled experts is missed by $CRPS$ and helps explain some of the downward scatter. 

\begin{figure*}[hbt!]
\centering
\includegraphics[width=0.65\linewidth]{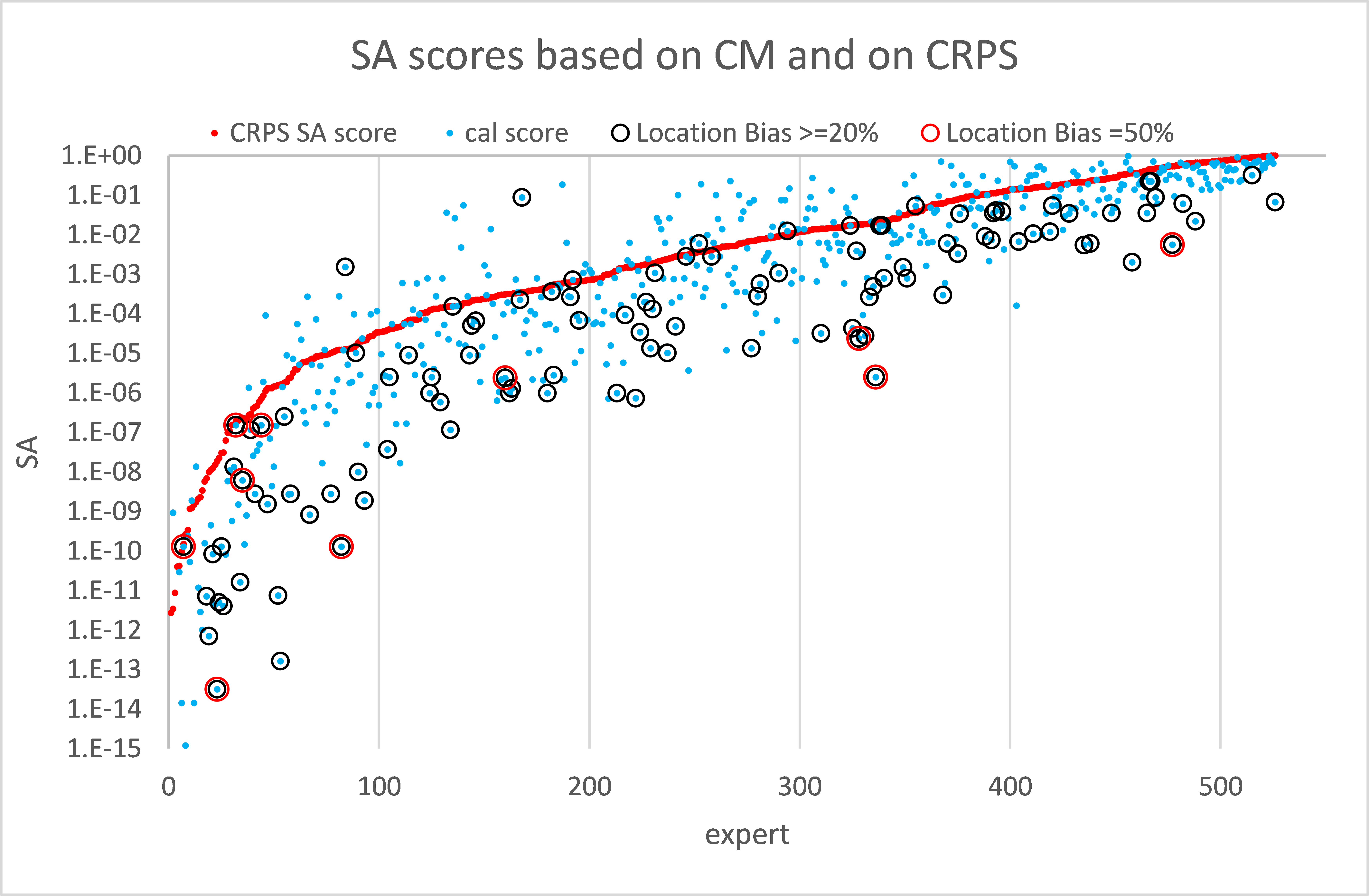}
\caption{Statistical accuracy of 526 experts with CM and CRPS with location biases circled. Location bias is the absolute difference between the percentage of medians above the realizations and $50\%$. Black circles denote $CM$ scores of those experts for whom the location bias is greater or equal to $20\%$. Red circles denote location bias of $50\%$.}\label{SA_CM_CRPS_LB}
\end{figure*}

For 70 experts, the location bias is 0\%.  These are termed experts without location bias (though of course there could be location bias in the lowest and highest inter-quantile intervals).  Figure \ref{SA_CM_CRPS_wo_LB} plots these $70$ $CM \,SA$ scores against all the $CRPS \,SA$ scores. On this subset, $CRPS$'s log geomean $SA$ score is $-2.50$ while that of $CM$ is $-2.68$. 

\begin{figure*}[hbt!]
\centering
\includegraphics[width=0.75\linewidth]{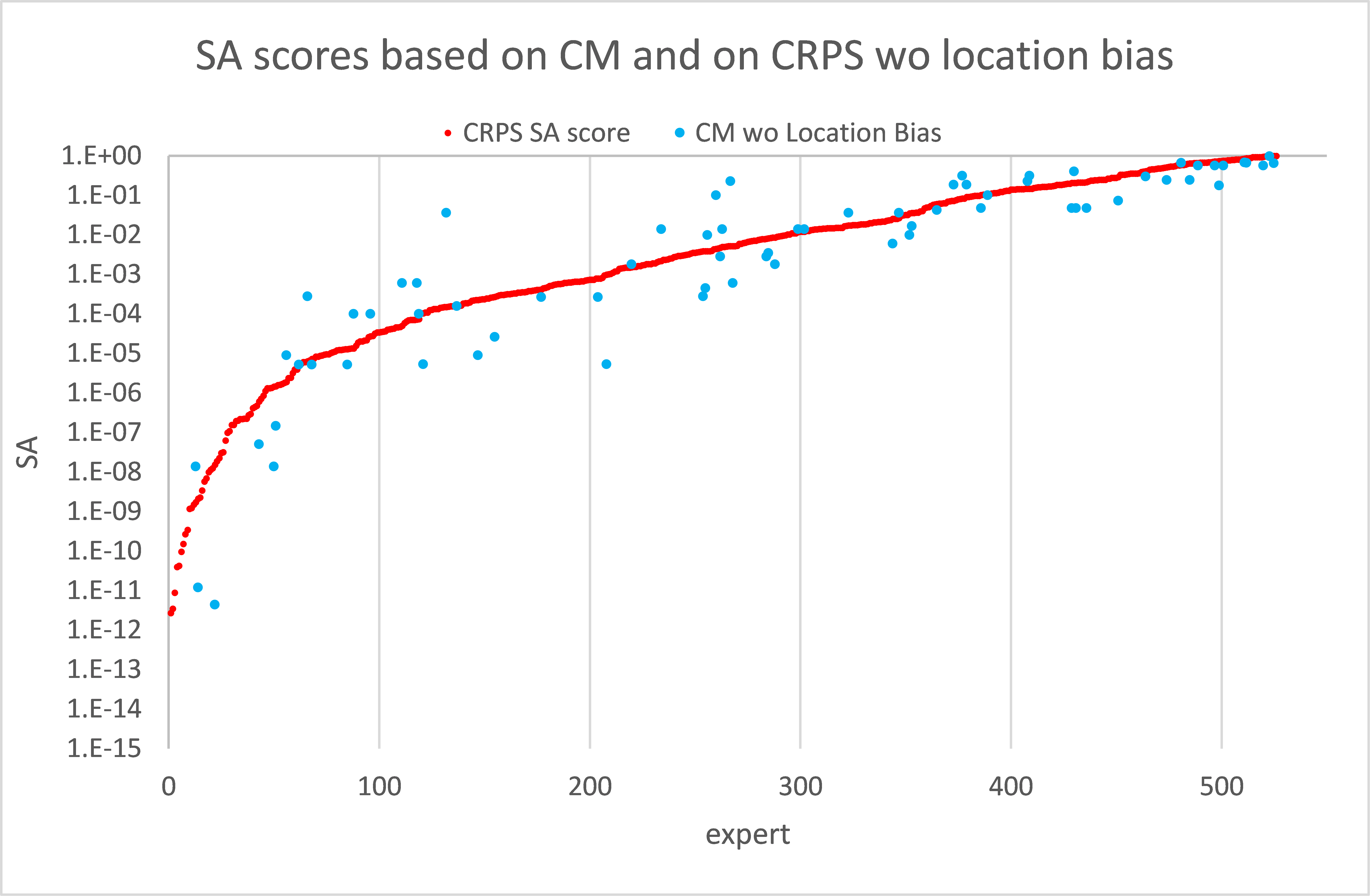}
\caption{Statistical accuracy of 526 experts with CRPS and 70 experts with CM without location bias.}\label{SA_CM_CRPS_wo_LB}
\end{figure*}

The number of calibration variables assessed in the 49 panels ranges from $7$ to $21$, see Figure \ref{NrCal}. This influences $CM\, SA$ in two ways, a larger number (i)  increases the accuracy of the $\chi^2$ approximation and (ii) tends to lower SA scores of poorly calibrated experts as the test of statistical accuracy has greater power. The first should decrease the differences between $CM\,SA$ and $CRPS\, SA$, at least for location unbiased experts, whereas the second enables a greater range of $CM\, SA$ scores and might therefore increase the differences. A multiple regression of $\log(\frac{CRPS\, SA}{CM\, SA}) $ against location bias and number of calibration variables explains $30\%$ of the variance (adjusted $R^2$) and both explanatory variables have a significant positive effect on the dependent variable. Thus, the influence of (ii) exceeds that of (i). The Pearson correlations of the dependent variable with location bias and with number of calibration variables are $0.41$ and $0.29$ respectively. The correlation of  the two explanatory variables is $-0.17$. A more detailed analysis might better explain the differences in the two $SA$ scores  but at this point it appears that location bias is the major factor.

\section{MAPE}
Statistical accuracy is not the only scoring variable of interest; the proximity of the median to the realizations is also important. There are many measures for such proximity \citep{morley2018measures, gneiting2007jrss}, each with benefits and  drawbacks. Perhaps the most popular is the Mean Absolute Percentage Error (MAPE), defined for forecasts $x_i$ and realization $r_i$  $\,i = 1 \dots n$ as
$$
\frac{1}{n} \sum_{i=1}^n|\frac{x_i-r_i}{r_i}|. 
$$
This is evidently unstable for very small $r_i$. Instability arises on this data set, as the largest $MAPE$ is over one million. The $MAPE$s of $326$ experts were less than $2$ and we focus on this subset. Figure \ref{MAPE} plots these $MAPE$ scores (left axis) and also plots the corresponding values of $CM\,SA$ and $CRSP\,SA$ (right axis). Although not overwhelmingly clear in the figure, the $CRSP\,SA$ scores tend to be higher than those of  $CM\,SA$, especially for very low $MAPE$s (see trend lines). The Spearman correlation of $CM\,SA$ and $MAPE$ on this data subset is $-0.15$ while that of $CRSP\,SA$ and $MAPE$ is $-0.26$. This results from the fact that $CRPS$ uses the (interpolated) $CDF$ whereas $CM$ is based on inter-quantile hit-rates. It is reasonable to expect that weighing experts according to $CRPS$ scores will produce better $MAPE$ values for the combination of experts than $CM$. Other researchers \citep{flandoli2011comparison}  have used likelihood scores based on interpolated $CDF$s and achieved better $MAPE$ performance than with $CM$, but such scores are notoriously improper. The great advantage of $CRPS$ in this regard is that it is based on a strictly proper scoring rule.
\begin{figure*}[hbt!]
\centering
\includegraphics[width=0.85\linewidth]{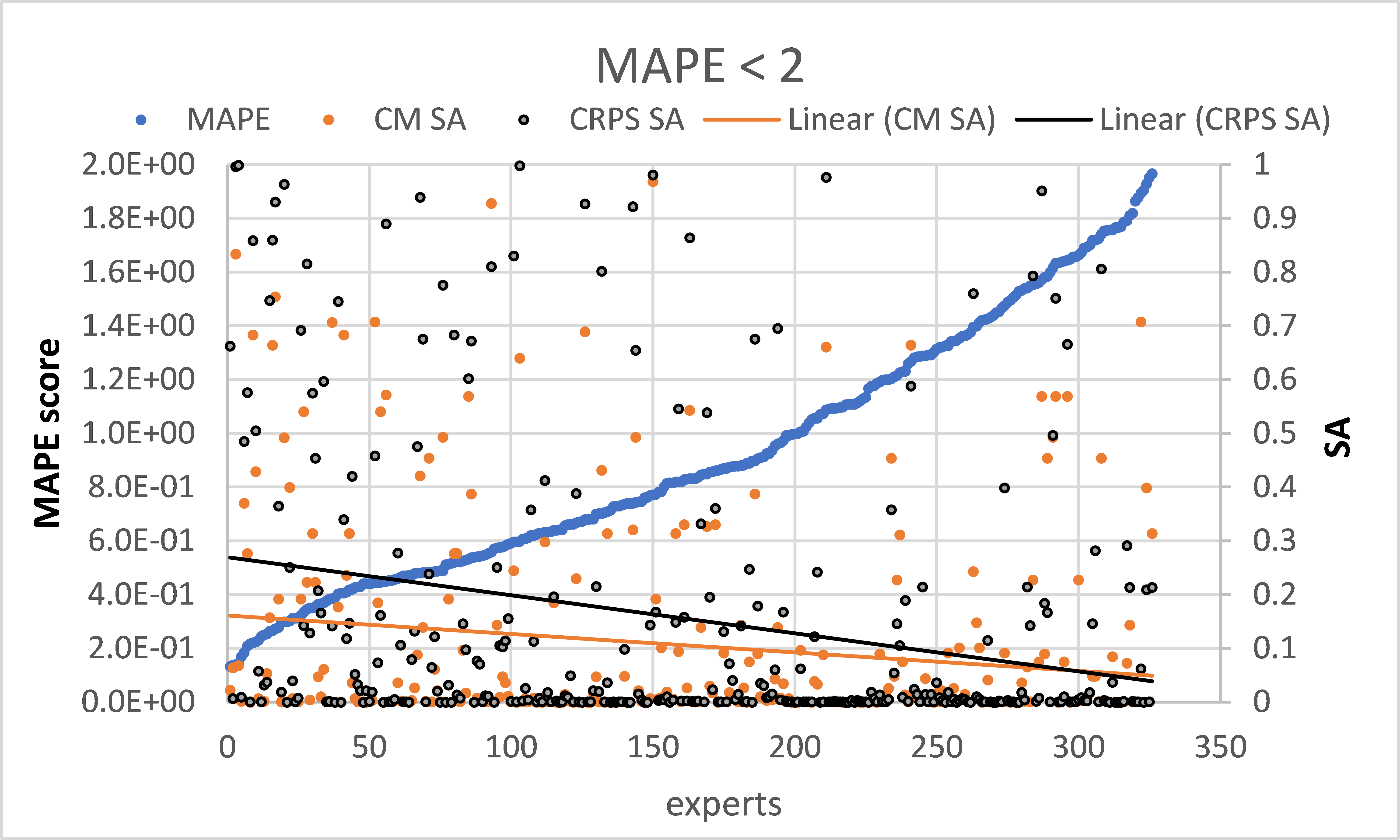}
\caption{326 MAPE scores $< 2$ (left axis) and SA for CM and CRPS (right axis), with trend lines.  }\label{MAPE}
\end{figure*}

\section{Conclusion}
A scale invariant version of the Continuous Ranked Probability Score (CRPS) has been developed and applied to an expert judgment data base involving $49$ studies with $526$ experts assessing in total $580$ calibration variables from their fields for which realizations are known. The transformed CRPSc yields a test for experts' statistical accuracy which has the advantage of a closed form solution without appeal to an asymptotic distribution. Compared to the statistical accuracy test used in the classical model it has the advantage of better rewarding proximity of a median point forecast to the realization. Nonetheless, the CRPS is insensitive to location and under-confidence bias. A future study will focus on combinations of experts' judgments, comparing the performance of $CRPS$  with other tests based on the Chi Square, the Kolmogorov Smirnov and the Cramer Von Mises statistics.

\bibliography{refs}
\end{document}